# JavaCtx: Seamless Toolchain Integration for Context-Oriented Programming.⋆

Guido Salvaneschi, Carlo Ghezzi, and Matteo Pradella

DEEPSE Group, DEI, Politecnico di Milano,
Piazza L. Da Vinci 32, Milano, Italy.
{salvaneschi,ghezzi,pradella}@elet.polimi.it

**Abstract.** Context-oriented programming is an emerging paradigm addressing at the language level the issue of dynamic software adaptation and modularization of context-specific concerns. In this paper we propose JavaCtx, a tool which employs coding conventions to generate the context-aware semantics for Java programs and subsequently weave it into the application. The contribution of JavaCtx is twofold: the design of a set of coding conventions which allow to write context-oriented software in plain Java and the concept of context-oriented semantics injection, which allows to introduce the context-aware semantics without a source-to-source compilations process which disrupts the structure of the code. Both these points allow to seamless integrate JavaCtx in the existing industrial-strength appliances and by far ease the development of context-oriented software.

## 1 Introduction

Context adaptation has been playing an increasingly important role in computer science, mainly because of emerging fields such as ubiquitous computing and autonomic computing. Ubiquitous computing [39] refers to anywhere and any time access to data and computing resources and requires mobile devices to adapt to continuously changing environmental and internal conditions such as connection bandwidth or battery power availability. Autonomic computing [27] refers to systems which overcome ever growing maintenance complexity through self-managing capabilities and require only high-level goal-oriented human guidance.

From a software engineering perspective, the design and the implementation of context-adaptable software is challenging, since adaptation must be performed at run time and context-specific behaviors typically crosscut the main modularization directions of the application. The problem has been tackled at different abstraction levels, and the proposed solutions encompasses approaches based on software architectures [31, 37, 32], middlewares [30], aspect-oriented programming [17] and computational reflection [36]. At the language level, context-oriented programming (COP) [25] was recently proposed as a solution specifically targeting the dynamic adaptation and modularization issues of context adaptation.

---

⋆ This research has been funded by the European Community's IDEAS-ERC Programme, Project 227977 (SMSCom).

Current COP implementations include a number of dynamic languages for which constructs leading context awareness can be implemented leveraging computational reflection [19]. COP has been introduced in statically typed languages with ContextJ [9], successively extended and currently one of the most mature COP implementations. ContextJ is an extension of the Java language with COP-specific constructs, such as a `with` statement that activates contextual adaptation in the dynamic extent of the scoped code block. Current COP Java extensions [9, 10, 26] employ source-to-source compiles which maps the contextual code to standard Java, rearranging the source structure in order to implement the context adaptation features.

Extending the language to introduce COP constructs leads to an elegant and essential syntax which perfectly fits the needs of context adaptation. However, the arising of great difficulties as a consequence of adding new language constructs is well known and has been referred as the problem of *tyrannical* constructs [16]. The issues introduced by language extensions, are especially relevant for those languages which require the addition of new keywords and source-to-source recompilation to be extended, since usually in the development process, beside compilers, a number of tools is employed in order to accomplish various support tasks and the problem of language compatibility with existing tools arises.

For example IDEs are usually aware of language syntax and semantics in that perform a whole of support tasks such as online incremental building, eager warning for compilation errors, syntax highlighting and content assist. Making them compatible with a language extension requires the development of plugins, which can be a non-trivial task in terms of both the initial effort and the post-release software maintenance if the reference quality levels are industrial-strength standards. The situation is even worse for those appliances such as test coverage suites of performance profilers which were not designed for extensibility and should be made aware of the new language. In the best scenario, since the result of their execution on the rearranged code produces meaningless results for the user, a tool-specific facility should be developed mapping the analysis results performed on the generated code back to the original one.

In this work we propose JavaCtx[1], an alternative solution to language extension and source-to-source transformation, employing the use of coding conventions and aspects to support context-oriented features. JavaCtx is a library and a tool that receives in input plain Java written according to a set of coding conventions and automatically generates the aspects and the classes implementing the contextual semantics. The aspects can be subsequently weaved using a standard aspect compiler and leaving the original code untouched. We refer to this phase as *contextual semantics injection*.

---

[1] Available at http://home.dei.polimi.it/salvaneschi/javactx

With JavaCtx, the programmer develops COP applications using plain and still semantically valid Java code. The injection of the contextual semantics is not disruptive in the sense that only intervenes in method dispatching, without deeply altering the code structure like a source-to-source compiler does. This allows to overcame the limitations above and makes the development of COP applications easier and seamless integrated with existing appliances.

The main contribution of this paper is the design and implementation of JavaCtx. The results we achieved in our work are the following:

- Introduction of the concept of contextual semantics injection, which allows COP as a programming methodology supported by an automatically generated aspect library rather than a language extension.
- Integration of context-oriented programming concepts with plain Java through a set of coding conventions and compatibility of the context-anabled code with existing development tools, such as IDEs, test coverage analyzers and performance profilers.
- Efficient implementation based on aspect-oriented programming with no additional overhead for non-contextual classes and methods.

The paper is organized as follows. In Section 2 we briefly introduce COP. In Section 3 we discuss the basic concepts of JavaCtx and present the details of its usage and implementation. Section 4 shows a general evaluation of the tool, encompassing integration with existing appliances and a performance evaluation. Section 5 discusses the related work. Section 6 draws some conclusions and presents future research.

## 2 Context Oriented Programming

COP is a programming paradigm oriented to run time adaptation and modularization of context-dependent behavior. The key concept of context-oriented programming is the *behavioral variation*. Behavioral variations define a unit of functionality which can be activated at run time and dynamically combined in order to produce the overall behavior of the program. A variation is referred to be *activated* when it starts affecting the software behavior. In COP variations activation is triggered by a context change. The notion of context adopted by COP is open and pragmatic: *context* is *any computationally accessible information* [14]. Interestingly, this definition does not limit to the stimuli that can come to the application from the external world, but encompasses the monitoring of the computing system itself as well.

The reification of behavioral variations in the programming language consists of *layers*, each one including a set of partial method definitions. For example the class `ResourceStorage` in Figure 1 implements a ContextJ object devoted to storing items which can be subsequently retrieved through a key. A cache and a logging functionalities can be dynamically activated. `ResourceStorage` contains two partial method definitions for the `request` method, one in the `caching` layer and one in the `logging` layer. A key feature of COP languages is that behavioral variations are first class citizens, i.e. layers can be assigned to variables, passed as functions parameter and returned as values.

The `with` statement is introduced by COP to activate one or more layers in the code block. As a precise design choice, layer activation is dynamically scoped. This fact has two implications. First of all layer activation obeys a stack-like discipline: a `with` statement *adds* one or more layers to the already existing ones and the previously active layer configuration is restored when the block scope expires. For example the `rs.request("Item3");` call is no more influenced by the activation of the `logging` layer in the previous `with` statement. Secondly, layer activation does not only affects the method calls directly performed in the syntactic scope, but also all those method executions triggered in turn in the control flow.

Partial method definitions are executed in reverse layer activation order. The combination among behavioral variations is obtained through the `proceed` call, which is similar to the `super` call in Java or to the `proceed` keyword in aspect-oriented programming. `proceed` dynamically dispatches the method call to the next active layer or to the original method if there are no more active layers. For example the call to the `rs.request("Item4");` method executes the partial definition in the `logging` layer which proceeds to the next active partial definition i.e. the one in the `caching` layer, finally proceeding to the base method.

The contextual language extensions proposed over the years interprets the COP paradigm according to the underlying programming model and provides language-specific functionalities. For this reason COP languages – though implementing the core concepts of the paradigm – come in a variety of different flavors. The features presented so far can be considered the basic functionalities of the object-oriented COP extensions. An overview of the more advanced constructs, for example the recent advances in event-driven context transitions is beyond the scope of this introduction. A fairly complete reference to the literature can be found in Section 5.

## 3  JavaCtx overview

In this section we present the key concepts of JavaCtx laying behind its design and implementation: contextual semantics injection, adoption of plain Java and source layout preservation. Hereafter we illustrate the coding conventions that are required to express the context-aware facilities in a JavaCtx application. Regarding this concern, our general approach in the design of the coding conventions was to keep the COP introduction as less intrusive as possible, while clearly indicate in the code where the contextual semantics comes into play. Finally we discuss some implementation details.

```
public Class ResourceStorage{          ResourceStorage rs =
                                           new ResourceStorage();
  layer caching {
    public String request(int req){    rs.request("Item1");
      System.out.println("Cache...")   with(logging){
      Object result = Cache.get(req);     rs.request("Item2");
      if (result == null){             }
        result = proceed(req);         rs.request("Item3");
        Cache.put(req,result);         with(caching,logging){
      }                                   rs.request("Item4");
      return result;                   }
  }}

  layer logging {                      EXECUTION:
    public String request(int req){
      System.out.println("Request!")   >Search
        return proceed(interaction);
  }}                                    >Request!
                                       >Search
  public String request(int req){
      System.out.println("Search")     >Search
      // retrieving item ...
      return item;                     >Request!
  }                                    >Cache...
}                                      >Search
```

**Fig. 1.** Partial method declaration and layer activation in context-oriented programming.

**Contextual Semantics Injection** The activation of behavioral variations requires to change the semantics of methods dispatching, since a method call must be rerouted according to the active layers and to the partial definitions declared in the target object. This result cannot be directly achieved adding a library to plain Java code. JavaCtx generates the machinery which implements the contextual semantics as a separate package containing aspects and class definitions. Finally the compilation with an aspect compiler waves the contextual semantics into the application (Figure 2).

**Plain and Valid Java Code** JavaCtx comes with a set of coding rules which allow to express COP constructs in plain Java. JavaCtx conventions include prescriptions for the names of partial method definitions, layers declarations and context-adaptable classes. At the cost of a small burden in the code this allows to take advantage of the context-adaptation facilities without recurring to a language extension.

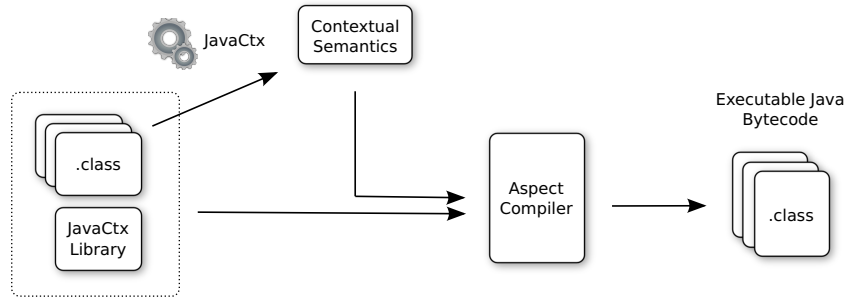

**Fig. 2.** The JAVACTX compilation process.

This design choice is a necessary premise for seamless integration with existing Java appliances. A direct, despite minor, advantage is to ensure IDE support without special-syntax aware plugins.

Even before the weaving of the contextual semantics, the code written according to the JAVACTX coding conventions is semantically valid and executable. The execution unfolds across only the basic version of each method and the contextual directives that activate the eventual behavioral variations are simply ignored.

**Code Layout Preservation** The weaving of the contextual semantics does not alter the code structure of the existing application. This feature makes JAVACTX different with respect to existing source-to-source approaches which map the code of the contextual application to a standard Java model, without taking any particular care of preserving its structure. For example a parameter referencing the current context is added to each method or method names are changed to redirect the original call to a synthetic proxy method.

In JAVACTX method calls are intercepted and rerouted according to the active layers. After rerouting, the actual method implementing the behavioral variation is executed, leaving its body and signature untouched. Interestingly, a local modification to the method body not involving contextually-dispatched calls does not even require to regenerate the contextual semantics.

Code layout preservation is central in the achievement of compatibility with existing appliances. The tools usually employed in support of the development process are still usable, thanks to the following facts: the context-enabled code is valid Java, and the original methods are effectively called with calls rerouted from *outside* the object (See Figure 3). In Section 4 we show an evaluation of how JAVACTX can be used in the Eclipse workbench with a code coverage analyzer and a performance profiler.

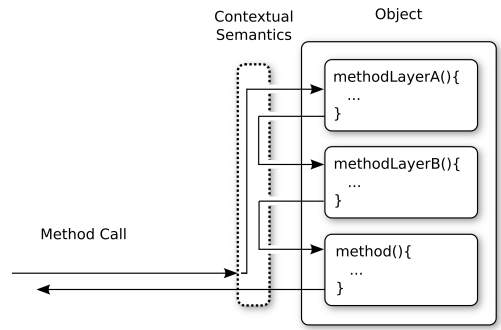

**Fig. 3.** Context-driven method dispaching in JAVACTX. Partial definitions are left untouched, while the call is rerouted from outside th object.

### 3.1 Using JavaCtx

**Layered Classes** The classes for which a contextual semantics has to be generated must be explicitly marked. This is accomplished by adding the `@LayeredClass` annotation, which informs the JAVACTX engine the the class contain partial methods definitions which requires context aware dispatching.

In order to keep the definitions of partial methods clean, we do not use the names of the layers including the whole package qualification, but the programmer defines local names for the layers referred in the definitions (see the next paragraph). The annotation parameters express the mapping from class-local layer names used in partial methods definitions inside the class to globally-scoped package-qualified layer names. For example in Figure 4 the class `Person` is declared as layered. Layers `myprj.layers.A`, `myprj.layers.B` and `myprj.inner.layers.C` are available in partial methods definitions respectively as `ALayer`, `BLayer` and `CLayer`.

**Partial Method Definitions** According to the requirement that JAVACTX code must be valid Java, each partial definition has a unique method identifier in the class. We adopted the code convention that partial method definitions declared in a certain layer have the name of the base method ending with the class-local name of the layer in which are declared. For example in Figure 4 three partial definitions are declared for the `print` method: in the `myprj.layers.A` layer, in the `myprj.layers.B` layer and in the `myprj.inner.layers.C` layer. JAVACTX automatically detects those methods for which a partial definition is declared, and generates the contextual semantics accordingly. Other methods, even inside a class marked `@LayedClass` are dispatched by the standard Java mechanism.

A fundamental point is how to express the `proceed` call, which is not a Java keyword. The JAVACTX coding convention is that a call to the base method in the syntactic scope of a partial method definition is dynamically dispatched across the remaining active layers. We point out that this convention does not constitute a limitation nor introduces any ambiguity in the code; instead it expresses the `proceed` directive in an elegant and clean way. *All* the invocations to the basic method in the syntactic scope of the partial definition are dynamically dispatched to the subsequent active layers, and therefore there is no possible ambiguity with a direct call to the basic method. On the other hand we argue that it is conceptually wrong to have a partial method that directly invokes the basic method without any contextual dispatching. Using a call to the base method to express `proceed` has also the consequence that since the signature of the dynamically chosen partial method definition must be the same of the base method, the type correctness of the `proceed` call is directly ensured by the Java compiler.

**Layer Declaration** Layers are standard Java classes extending the `Layer` class provided by the JAVACTX library. The only convention regarding layers declaration is that the class must declare an `id` field, marked with the `public`, `static` and `final` modifiers:

```
import cop.Layer

public class A extends Layer {
    public static final id = new A();
}
```

This allows to refer to the layer `A` with the compact notation `A.id` and to treat layers as singleton objects without further burden. The convention is due to the precise design choice to enforce type coherence among COP entities. Layers compare in the code as objects of type `Layer`, which is a natural approach in an OO system. The option of using a notation like `A.class` adopted by other Java COP extensions [11], instead, leads to the counterintuitive situation that layers appear in the code as objects of type `class` instead of type `Layer`.

**Layer Activation** Layers are activated by the invocation of the `withActiveLayers` static method on the `Ctx` class, which exposes the JAVACTX API and is the entry point for accessing all the context-related facilities. Like the `Layer` class, `Ctx` is part of the `cop` package provided by JAVACTX.

```
Cxt.withActiveLayers( layers );
  ... // Adapted block
Ctx.end();
```

In line with other COP languages, the layer to activate can be known only at run time, which is especially useful in case of external, sensor based context providers, which dynamically decides layer activation. The resulting code is structured like `withActiveLayer(contextProvider.getActiveLayers())`. `Ctx` implements different flavors of the `withActiveLayers(...)` method, which for example can accept a `Collection` of layers to activate. The other basic constructs of COP implementations, such as the `Ctx.withoutLayers` method are available as well.

The coding conventions require that the programmer calling a directive to the `Ctx` class is responsible to assure that it is followed by a `Ctx.end()` call *in the scope of the same codeblock*. This convention is necessary in order to syntactically delimit the effect of a layer activation and enforce the dynamic extent activation policy. Despite this being probably the more obliging of the rules required by JavaCtx, it is comparable (yet simpler) to code conventions imposed by widespread programming frameworks, e.g. conventions for memory allocation in Cocoa [1]. Moreover, due to its purely syntactic nature, static checking is straightforward.

As already mentioned, layer activation have consequences only after contextual semantics injection. If the contextual semantics has not been weaved yet, the calls to the `Ctx` class simply have no effect.

### 3.2 Implementation Details

We implemented JavaCtx as a source analyzer and code generator written in Java. The tool leverages Java annotations processing to detect those classes in the classpath that require to be augmented with the contextual semantics. Java reflection is used to detect Layer declarations and partial method definitions. For the code generation phase we employed Apache Velocity [3], a specification language and fast template rendering engine. The templates are parametrized with reference objects defined in the Java code; after the information for reference instantiation is collected from the codebase, the parameters are filled with the actual values and the templates are instantiated. The generated code includes standard Java classes and aspects, which are weaved by the AspectJ 5 compiler.

For what concerns the aspects generation, the design of JavaCtx was lead by the constraint of using only statically determinable join points to avoid incurring in performance penalties due to runtime conditions evaluation.

A final remark regards concurrency. Since COP dynamic activation requires *per control flow* variation activation and each thread must adapt independently, the generated code internals employ `ThreadLocal` references for active layers.

## 4   Evaluation

We evaluated the integration of JavaCtx in the development toolchain experimenting its usage in the Eclipse workbench combined with the Eclemma [5] and the YourKit [7] plugins. For what concerns the user experience with Eclipse, it does not degrade with JavaCtx, and IDE functionalities, e.g., content assist, coding-time errors warning and continuous compilation, work correctly.

```
@LayeredClass({
    "ALayer => myprj.layers.A",
    "BLayer => myprj.layers.B",
    "CLayer => myprj.inner.layers.C"
})
public class Person {

    public String printALayer(String s){
        String r = print(s);
        return "A" + r;
    }

    public String printBLayer(String s){
        String r = print(s);
        return "B" + r;
    }

    public String printCLayer(String s){
        String r = print(s);
        return "C" + r;
    }

    public String print(String s){
        return r + "Base";
    }
}
```

**Fig. 4.** Layered class declaration and partial method definitions in JavaCtx.

Eclemma is a Java code coverage appliance for the Eclipse platform, internally based on the EMMA [2] code coverage tool. Eclemma adds an execution mode to the Eclipse projects or unit tests, which collects coverage information. The results are both available in the form of analysis reports and source highlighting. We setup a project executing all the methods of each class and with all the partial methods definition *proceed* up to the base method. We ran Eclemma both before and after the contextual semantics injection, activating with an initial `withActiveLayers` call, every available layer. As expected, before the injection only the base methods were marked as executed, while after the injection Eclemma correctly highlights also all the partial definitions.

YourKit is a commercial Java performance profiler. Among the other analysis features such as memory leaks detection, it shows the time spent in the execution of each method of a call stack. In this case the use of JavaCtx resulted not completely transparent, since the calls performed in the machinery implementing the contextual dispatching, such as the aspects intercepting the original call, are displayed to the user. However this does not invalidate nor alter the clearness of the analysis since after the spurious calls due to aspect weaving, either the partial definition or the original method is correctly reported with the estimated performance.

### 4.1 Performance

According to Gregor Kiczales, one of the fathers of AOP, aspects are the "15% solution", in the sense that you will use it for about 15% of a typical project [4]. We argue that for real-world sized projects a similar consideration holds also for COP. Therefore it is of great importance that the natural performance overhead introduced by COP with context-driven dispatching not only is as low as possible, but is also limited to the portion of code actually context-aware, without performance impacts on the rest of the codebase. This was one of the cornerstone design requirements for JavaCtx. As a result non context-aware method calls are not influenced by the contextual semantics and do not incur in any performance penalty.

We experimented JavaCtx[2] the performance of JavaCtx calling a contextually dispatched method for which five partial definitions exist in different layers. All the partial definitions execute a `proceed` statement. We evaluated the time spend by $10^7$ calls to the method varying the number of active layers (and therefore of involved partial definitions) both for JavaCtx and ContextJ. We compared this results with a similar setup in plain Java, with six methods each calling the next one up to a base one and changing the method initially called. The choice of comparing JavaCtx against ContextJ is that it is the fastest COP Java implementation available and currently one of the fastest COP languages [8, 38].

To allow the JVM to perform optimizations, each measure was taken with a dry run before the real test. The results are reported in Figure 5. Not surprisingly the performance are rather worse than ContextJ which can rely on code rearrangement and does not pay the overhead of aspects interception. However even with a test which focuses on pure method dispatching, which is the part of the execution that more intensively is slowed down by context awareness, the performance results are comparable. We ran our experiments on a machine equipped with an Intel Core 2 Quad Q9550, 4GB RAM, Windows XP SP2, Java 6 and AspectJ 5.

### 4.2 Current Limitations

Many COP languages support *before* and *after* methods. The current version of JavaCtx supports only around methods, but could be easily extended. Clearly this is not a serious limitation, as around methods are the most general form. However using the simplest form appropriate for the task more clearly expresses the programmer's intent.

---

[2] The code used for the performance experiments is available at the website: http://home.dei.polimi.it/salvaneschi/javactx

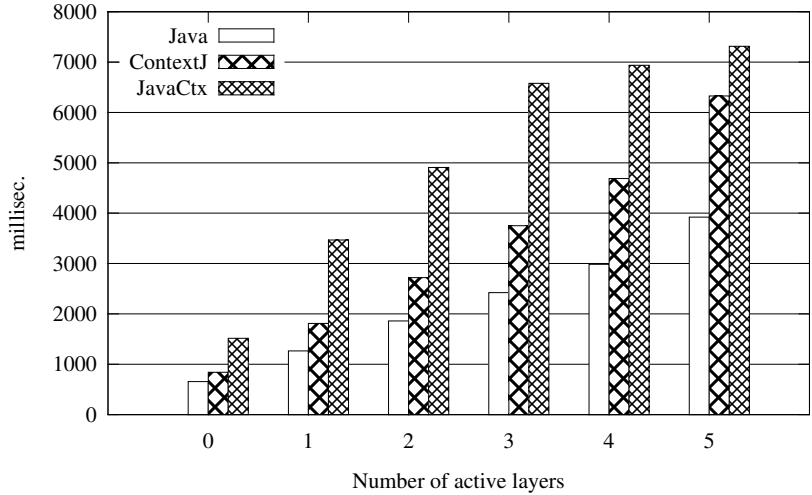

**Fig. 5.** Performance evaluation of JavaCtx: time required to execute $10^7$ method calls in function of the number of active layers.

Another current limitation of JavaCtx is the lack of a reflection API allowing to inspect active layers and possibly to reflectively change the current layer configuration. While these features were already explored in literature [10], it is not clear to us how powerful such an API should be in terms of direct access to layer activation in order to be both useful and not potentially unsafe. For this reason we believe that such advanced functionalities deserve further reasoning and we left the implementation of such API for the future after deeper experimenting with applications.

## 5 Related work

Context-oriented programming have been firstly proposed in the pioneering work of Costanza [14, 13, 15] on ContextL, an extension of Common Lisp based on the CLOS metaobject protocol [29]. Over the years several COP extensions of existing languages have been proposed, such as Python, Ruby, Smalltalk, Scheme, Javascript and others. A fairly complete review of these works with a performance comparison can be found in [8]. Ambience [23], originally proposed simultaneously with ContextL, is a context-aware object system for Lisp based on prototypes and multimethods for context-driven dispatching.

Statically typed languages with limited reflective capabilities are less flexible, so it is harder to define COP extensions for them. ContextJ* was a first attempt, which tried to directly reproduce the COP extensions in Java [25], but exhibiting limited performances and a rather complex syntax. ContextLogicAJ [11] was a prototype based that lead to the development of ContextJ [9]. The approach used in JavaCtx has similarity to the one of ContextLogicAJ [11], since both are based on aspects. However to the best of our knowledge, ContextLogicAJ has never been considered more than the first prototype for ContextJ and has never been released. ContextJ was the first Java COP extension which actually implemented the COP constructs through a source-to-source compiler. Interestingly, it offers a rich reflection API which allows to inspect the contextual entities and their current activation state. JCOP [10] is an extension of ContextJ that supports declarative variation activation defined on an enabling condition, in a way similar to conditional pointcuts in AOP. This feature reduces the need for the developer to specify variation activation explicitly through `with` statements.

EventCJ [26, 38] is a Java-based COP language whose design was driven by motivations similar to JCOP's: context change triggered by events. EventCJ allows to declare context transition rules: the firing of an event can trigger the automatic (de)activation of a set of layers. Beside the usual per control flow activation mechanism, EventCJ pushes the adaptation capabilities of COP applications to the object level, in that layer activation can be triggered independently on each single object.

ContextErlang [21, 22] is a context-oriented version of the Erlang language, based on the integration of the actor concurrency model, originally proposed by Hewitt [24], with the COP paradigm. ContextErlang agents exchange special context-change messages that trigger adaptation on the other agents. Therefore some features which require ad-hoc techniques in thread-based languages, such as event driven context switch or asynchronous variations activation, are managed automatically.

The context-oriented paradigm has some similarities with AOP. Both propose a solution that allows the programmer to cope with the issue of cross-cutting concerns. However while AOP mainly focuses on effective code modularization, COP focuses on dynamic activation and composition of behavioral variations. Dynamic AOP shares with COP the idea of run time activation of a code unit implementing some variation to the basic flow of the application. Dynamic AOP have been implemented in several research frameworks such as Prose [34], JAC [33] and AspectWerkz [12]. AspectJ [28] has some dynamic activation facilities as well. For example the `percflow` pointcut allows to execute an advice only if another pointcut is in the given control flow and the `target` pointcut expresses a condition on the run time of the target of the method call. However not all of these features are available in other industrial-strength AOP frameworks. For example the Spring AOP pointcut model [6] does not support `percflow` pointcuts. The main difference between these technologies and COP is that they generally address the issue of run time activation of crosscutting functionalities, while COP introduces language constructs specifically designed to tackle the problem of context adaptation.

Event-based programming [18] is a programming paradigm specifically introduced for managing reactive behavior. It relies on the idea of implicit invo-cation [20] and introduces language-level features to express events and notifications. Event-based languages differ form contextual languages in that they directly trigger *actions* rather than layers recombination and behavioral adaptation. However advanced features of event-based languages can be used to implement a form of behavior combination similar to AOP. For example Ptolemy [35] allows to associate closures to events, that are subsequently executed by the event receiver. Therefore the overall behavior of the application is a dynamic combination of the behaviors of the caller and of the callee.

## 6    Conclusions and Future Work

In this paper we presented JavaCtx, a Java COP extension specifically aimed at easing the development of context-aware applications through the concept of contextual semantics injection, that allows a seamless integration with existing tools. Our future research directions are the following. First, we plan to strengthen the JavaCtx set of available features, implementing at least part the advanced functionalities, for example context-specific reflection.

Second, we intend to take advantage of JavaCtx for a considerable effort in the development of context-aware applications, in the wide scope of the ERC project SMSCom (*Self Managing Situated Computing*) we are participating in, with the aim to evaluate the COP paradigm in real-world middle-sized projects. *Situational* indicates that software behaves according to the evolving situation in which it operates, in the case of COP this is clearly provided by the *context*. Developing and running situational software imposes a paradigmatic shift from a conventional development, to new scenario in which bits of applications are composed in possibly unpredictable ways. COP appears to be a natural approach for developing this kind of applications.

Lastly, we plan to use JavaCtx as a testbed for experimenting new constructs that allow context adaptation to be more effective and easy to implement.